# Kelvin-Helmholtz instability of the collisionless anisotropic space plasma

N. S. Dzhalilov,[1]⋆ R. Ismayilli,[2]
[1]*Shamakhy Astrophysical Observatory, Shamakhy, Azerbaijan*
[2]*Center for Mathematical Plasma-Astrophysics, KU Leuven, Leuven, Belgium*

**ABSTRACT**
The linear MHD Kelvin-Helmholtz instability (KHI) in an anisotropic plasma concerning the direction of an external magnetic field is examined in detail. For this purpose, the MHD equations are used to describe the motion of plasma as a fluid, which is derived from 16 moments of Boltzmann-Vlasov kinetic equations for collisionless plasma. In addition, the heat flux along the magnetic field is taken into account. The growing rates of KHI are calculated as functions of the anisotropic plasma properties for a shear flow along the magnetic field at supersonic velocities. On the other hand, the quasi-transverse propagation of surface waves between flows with varying velocities is thoroughly examined for both zero-width and finite-width transition layers. In contrast to the tangential discontinuity, it is proved that the limiting breadth of the transition layer constrains the KHI excitation as the wavenumber grows. The instability under investigation could be one of the main ways of dissipation of large-scale low-frequency Alfén wave turbulence existing in the solar wind plasma.

**Key words:** Anisotropic plasma – MHD shear instability – Solar wind

## 1 INTRODUCTION

The evolution of shear instabilities of the Kelvin-Helmholtz type (KHI) in fluid flows with a transverse velocity gradient is one of the widespread phenomena in space plasmas. The simplest but effective method for describing this instability is to consider two homogeneous flows with different velocities and the tangential discontinuity between them. Pioneering theoretical studies on these issues analyze the instability of shearing hydrodynamic subsonic flows (e.g. Fejer 1964; Gerwin 1968; Southwood 1968; Dobrowolny 1972). In addition, nearly all the books on fluid mechanics cover the fundamentals of this instability theory (e.g. Lamb 1945; Chandrasekhar 1961; Turner 1973).

Numerous detailed studies compared with other types of instabilities in fluids are related to the fact that KHI is considered one of the primary mechanisms for the transfer of energy and momentum from one medium to another through the generation of turbulence via cascades of interacting vortices in the contact surface between them. For example, the transfer of energy and momentum from the solar wind to the magnetosphere of planets using surface waves has been extensively discussed both theoretically and in data analyses (Southwood 1974; Chen & Hasegawa 1974; Scarf et al. 1981; Miura 1984, 1992; Thomas & Winske 1993; Keller & Lysak 1999; Barranco 2009). Many theoretical works are devoted to studies of the emergence of this type of large-scale instabilities in the conditions of cometary tails (Ershkovich et al. 1972; Min 1997; Downes & Ray 1998; Ershkovich & Mendis 1983; Ershkovich et al. 1986), in the interstellar medium, accretion discs, and supernova remnants (Shen et al. 2006; Barranco 2009; Wang & Chevalier 2001), also, in CIRs of solar wind (Korzhov et al. 1984), at the heliopause, where the solar wind is halted interacting with interstellar medium (Chalov 1996), at the magnetopause (Hasegawa 1975), Earth's aurora (Farrugia et al. 1994), protoplanetary disks [(Gómez & Ostriker 2005), jets and outflows in subsonic, transonic, supersonic and relativistic regimes (Turland & Scheuer 1976; Blandford & Pringle 1976; Ferrari et al. 1980; Keppens et al. 1999; Baty & Keppens 2006) and among many other situations. Many theoretical papers on the KHI in solar plasmas, (e.g. Karpen et al. 1993; Andries & Goossens 2001; Soler et al. 2010; Zaqarashvili et al. 2010; Díaz et al. 2011) has been published.

It is well-founded that instability arises when the velocity gradient or jump between the velocities in the flows exceeds some threshold value (Chandrasekhar 1961; Talwar 1964; Ferrari et al. 1981; Pu 1989). This threshold also depends on the physical parameters of the steady flows. In the absence of a magnetic field in incompressible flows, instability is possible for any shifts in velocity. The presence of a magnetic field along the flow suppresses the instability. For example, the suppression of the KHI of fully ionized shear sub-Alfvénic plasma flows by a magnetic field is considered in (Chandrasekhar 1961; Elphic & Ershkovich 1984). If the plasma is partly ionized, the issue becomes more complex., e.g., (Soler et al. 2012).

Linear, primarily analytical analyzes for a tangential discontinuity show that the growth rate of the KHI is proportional to the wavenumber. It means that the growth rate is significant for short waves. However, strictly speaking, the assumption of a tangential discontinuity between the flows with relative motions is valid only for long wavelengths and breaks down for short ones. Consequently, many studies of the KHI are based on the models in which the finite-width transition layer is assumed as a vortex sheet of the shear flow velocity profile (Michalke 1964; McKenzie 1970; Blumen et al. 1975; Drazin & Davey 1977; Miura & Pritchett 1982; Roychoudhury & Lovelace 1986; Ray & Ershkovich 1983; Miura 1984; Roy Choudhury 1986b;

⋆ E-mail: namigd@mail.ru

Amerstorfer et al. 2007). Most parts of these works are devoted to the shear flow instability in an unbounded plasma. Effects of bounded plasma flow to KHI are also studied in, e.g., (Leonovich & Mishin 2005).

The numerical MHD simulations of evolutions of the KHI (e.g., Miura 1982, 1984, 1987, 1997, 1999; Wu 1986; Manuel & Samson 1993; Thomas & Winske 1993; Otto & Fairfield 2000; Brackbill & Knoll 2001) predict that fast growth of KH waves (or vortices) with smaller wavelengths (or length scales) results in the early their saturation. So, the KHI in its nonlinear stage can develop the small-scale filamentary field and current structures. Inside these narrow current layers can occur the magnetic reconnection, resulting in plasma transport (Miura 1992). However, for the smallest length scales close to the ion inertia scale, the ideal MHD approximation is not valid anymore, and the Hall term in Ohm's law must be included (Huba 1994; Nykyri & Otto 2004). In work (Fujimoto & Terasawa 1991), authors have carried out the study of the ion inertia effect on the KHI of two-fluid plasma and in (Wolff et al. 1980; Huba 1996) included finite Larmour radius (FLR) effects in the incompressible MHD equations. Furthermore, they discussed the effect of a finite boundary thickness on the KHI and mentioned that both FLR and finite thickness effects stabilize the flow at short wavelengths. Many works have been done on numerical simulations of KH instabilities, considering the resulting magnetic field generation (Alves et al. 2012; Srinivasan et al. 2012; Modica et al. 2013; Modestov et al. 2014).

In all the works mentioned above, KHI studies were carried out based on isotropic MHD equations, valid for a collisional plasma. In most cases, however, when KHI can occur and is observed in cosmic situations, the plasma is weakly collisional or collisionless (for example, solar and stellar wind, heliosphere, planetary magnetosphere, cosmic plasma jets, Etc.). Under such conditions, the plasma becomes anisotropic concerning the direction of the magnetic field. The KHI has been discussed in anisotropic plasma using Chew, Goldberger, and Low (CGL) equations (Chew et al. 1956) for the situations where collisions are not sufficiently strong to keep the pressure a scalar but sufficiently strong prevent the heat flow and other transport processes. Authors (e.g Talwar 1965; Roy Choudhury 1986a; Brown 1999; Ramos 2003; Modica et al. 2013) have discussed the MHD KHI problem in different aspects using the CGL fluid approximation. The KHI of an anisotropic, finite-width, supersonic shear layer and the nonlocal coupling of the firehose and mirror instabilities via a spatially varying velocity are investigated in (Roy Choudhury & Patel 1985).

The main disadvantage of the CGL fluid approximation for anisotropic cosmic plasmas is that two unjustified polytrope laws are introduced and do not consider the anisotropic heat flux that occurs naturally in the presence of a strong external magnetic field. These disadvantages are eliminated in the MHD transport equations of anisotropic plasma, which are derived as 16 integral moments of the Boltzmann kinetic equation (Oraevskii et al. 1985; Ramos 2003). These equations take into account the heat flux along the magnetic field. In addition, these equations allowed (Ismayilli et al. 2016a,b; Uchava et al. 2020) to address some difficulties regarding the emergence of KHI in anisotropic plasma with the shear flow.

In this paper, we consider the occurrence of KHI in an anisotropic plasma based on 16-moment MHD equations taking into account the heat flux along the magnetic field. First, a general disturbance equation is derived for the linear stage of instability, where the steady flow velocity has an arbitrary profile. The KHI is then discussed for cases of tangential discontinuity and transition zone with a finite width between flows. Finally, growth rates are calculated for solar wind CIR parameters depending on the transition layer width.

## 2 GOVERNING EQUATIONS

The 16-moment set of equations may be used to fluid describe a collisionless anisotropic plasma, which is comprehensive because it includes the evolution of heat fluxes along the magnetic field (Dzhalilov et al. 2008, 2011; Ismayilli et al. 2018). Furthermore, these equations apply to plasmas with a single component (ion):

$$\frac{d\rho}{dt} + \rho \operatorname{div} \vec{V} = 0, \quad (2.1)$$

$$\rho \frac{d\vec{V}}{dt} + \nabla \left( p_\perp + \frac{B^2}{8\pi} \right) - \frac{1}{4\pi} (\vec{B} \cdot \nabla) \vec{B} = \vec{h}_B \left( \vec{h}_B \cdot \nabla \right) \left( p_\perp - p_\parallel \right) \\ + \rho \vec{g} + \left( p_\perp - p_\parallel \right) \left[ \vec{h}_B \operatorname{div} \vec{h}_B + \left( \vec{h}_B \cdot \nabla \right) \vec{h}_B \right], \quad (2.2)$$

$$\frac{d}{dt} \frac{p_\parallel B^2}{\rho^3} = -\frac{B^2}{\rho^3} \left[ B \left( \vec{h}_B \cdot \nabla \right) \left( \frac{S_\parallel}{B} \right) + \frac{2S_\perp}{B} \left( \vec{h}_B \cdot \nabla \right) B \right], \quad (2.3)$$

$$\frac{d}{dt} \frac{p_\perp}{\rho B} = -\frac{B}{\rho} \left( \vec{h}_B \cdot \nabla \right) \left( \frac{S_\perp}{B^2} \right), \quad (2.4)$$

$$\frac{d}{dt} \frac{S_\parallel B^3}{\rho^4} = -\frac{3p_\parallel B^3}{\rho^4} \left( \vec{h}_B \cdot \nabla \right) \left( \frac{p_\parallel}{\rho} \right), \quad (2.5)$$

$$\frac{d}{dt} \frac{S_\perp}{\rho^2} = -\frac{p_\parallel}{\rho^2} \left[ \left( \vec{h}_B \cdot \nabla \right) \left( \frac{p_\perp}{\rho} \right) + \frac{p_\perp}{\rho} \frac{p_\perp - p_\parallel}{p_\parallel B} \left( \vec{h}_B \cdot \nabla \right) B \right], \quad (2.6)$$

$$\frac{d\vec{B}}{dt} + \vec{B} \operatorname{div} \vec{V} - (\vec{B} \cdot \nabla) \vec{V} = 0, \quad (2.7)$$

$$\operatorname{div} \vec{B} = 0, \quad (2.8)$$

where $\rho$ is the density, $p_\parallel$ and $p_\perp$ are the parallel and perpendicular gas pressures, $\vec{B}$ is the magnetic field, $\vec{V}$ is the bulk velocity of the plasma, $\vec{g}$ is gravitational acceleration, $\vec{h}_B = \vec{B}/B$ and $d/dt = \partial/\partial t + (\vec{V} \cdot \nabla)$ is the convective derivative. Here, $S_\parallel$ and $S_\perp$ are the heat fluxes along the magnetic field caused by ions moving in parallel and perpendicular thermal kinetic motions, respectively. If heat fluxes are ignored, that is, when $S_\parallel = 0$ and $S_\perp = 0$, we obtain a closed system of equations known as the CGL (Chew-Goldberger-Low) equations via Eqs. (2.1–2.4), (2.7), and (2.8); see the seminal work by Chew et al. (1956). The heat fluxes can be incorporated into the 16-moment set of Eqs. (2.1-2.8) to give a more comprehensive form than the CGL equations.

In order to analyze the implications of these equations to KHI problem, we assume that in the background circumstances, the plasma is homogeneous, $g = 0$, and $\rho_0, p_{\perp 0}, p_{\parallel 0}, B_0, S_{\perp 0}, S_{\parallel 0} = const$. The magnetic field is directed along the $z$-axis in the Cartesian coordinate system, and the flow velocity directed along the magnetic field has a transverse gradient, $V_0 = V_0(x)$. In Ismayilli et al. (2018) paper, we derived the wave equation in general form for linear perturbations of physical quantities, which is $\sim Y(x) \exp i(k_y y + k_z z - \omega t)$ ($\omega$ - oscillation frequency, $k_y, k_z$ - wavenumbers):

$$\frac{d}{dx} \left( A(x) \frac{dY(x)}{dx} \right) - k^2 \beta_A(x) Y(x) = 0, \quad (2.9)$$

where $Y(x)$ is perturbed amplitude, wave numbers are $k^2 = k_y^2 + k_z^2$,

$k_y^2 = k^2(1-l)$, and $k_z^2 = k^2 l$, $l = \cos^2(\theta)$, $\theta$ - wave propagation angle relatively to magnetic field directed on the z-axis. Here

$$A(x) = \frac{\beta_A}{\chi^2}, \quad \chi = \left(1 - l + l\frac{\beta_A}{\beta_*}\right)^{1/2},$$

$$\beta_A = \beta + \alpha - 1 - \xi^2,$$

$$\beta_* = \beta + 2\alpha + 2\alpha^2 \frac{\left(\xi^4 + 2\gamma\xi^3 + 2\gamma^2\xi^2 - 5\xi^2 - 6\gamma\xi + 3\right)}{\left(\xi^4 - 6\xi^2 - 4\gamma\xi + 3\right)\left(\xi^2 - 1\right)}, \quad (2.10)$$

$$\xi(x) = \frac{\omega - k_z V_0(x)}{c_\parallel k_z},$$

$$\alpha = \frac{p_\perp}{p_\parallel}, \quad \beta = \frac{B^2}{4\pi p_\parallel} = \frac{v_A^2}{c_\parallel^2}, \quad \gamma_\perp = \frac{S_\perp}{p_\perp c_\parallel}, \quad \gamma_\parallel = \frac{S_\parallel}{p_\parallel c_\parallel},$$

where $v_A$ is the alfvenic velocity and $c_\parallel$ is a parallel sound speed. Here plasma anisotropic parameter $\alpha$, magnetic parameter $\beta$ and heat flux parameters of $\gamma$ are defined by initial unperturbed plasma parameters. Consider the case of two uniform flows with different velocities $V_{01}$ and $V_{02}$ separated by a contact layer with a thickness of $2L$ (Fig. 1). Let $\overline{V} = \frac{1}{2}(V_{01} + V_{02})$ mean velocity, $h = \frac{V_{01}}{V_{02}} \geq 1$ jump index in velocity, $M = \frac{\overline{V}}{c_\parallel}$ Mach number, $V(x) = \frac{V_0(x)}{\overline{V}}$ normed velocity. As we have shown in Ismayilli et al. (2018) resonant interaction of the wave with the flow occurs if $\omega \approx k_z \overline{V}$. So let $\omega = k_z \overline{V}(1 + \Omega)$ and $\Omega$ is unknown complex spectral parameter. Then $\xi(x) = M(\Omega + 1 - V(x))$.

Instability growth rate is defined as

$$\Gamma = \frac{\operatorname{Im}(\omega)}{\operatorname{Re}(\omega)} = \frac{\operatorname{Im}(\Omega)}{1 + \operatorname{Re}(\Omega)}. \quad (2.11)$$

We will consider the model problem of a shear layer with a finite width $2L$, which arises between fast and slow flows, Fig. 1. Depending on the flow velocity, as well as on the physical parameters of the plasma in these flows, the width of the transition region with a gradient can be very narrow in comparison with the wavelength (long-wavelength approximation, $\lambda \gg L$), large (short-wavelength approximation, $\lambda \ll L$), or the same of order ($\lambda \sim L$). To find out how the width of the gradient layer can affect the condition for the occurrence of KHI, it is necessary to solve eq. (2.9) under the given boundary conditions. It is not possible to solve this equation in the general case when the velocity profile with a gradient is not specified. Below we will consider three special cases: tangential discontinuity, linear and hyperbolic velocity profiles.

## 3 TANGENTIAL DISCONTINUITY

With a very narrow layer, when $L \to 0$, a tangential discontinuity arises between the flows (dotted curve in Fig. 1). In this case the velocity profile may be described by the Heaviside step function of

$$V_0(x) = \begin{cases} V_{01}, & x < 0 \\ \overline{V}, & x = 0 \\ V_{02}, & x > 0 \end{cases}$$

The dispersion equation for $\Omega$ was derived in Ismayilli et al. (2018) by applying boundary conditions at the discontinuity. In the case when all physical parameters in both streams are the same, except for the velocities, and therefore, $V_0 = V_{01}, \xi = \xi_1 = M(\Omega - \Delta)$ at $x < 0; V_0 = V_{02}, \xi = \xi_2 = M(\Omega + \Delta)$ at $x > 0$ and $V_0 = \overline{V}, \xi = \xi_0 = M\Omega$

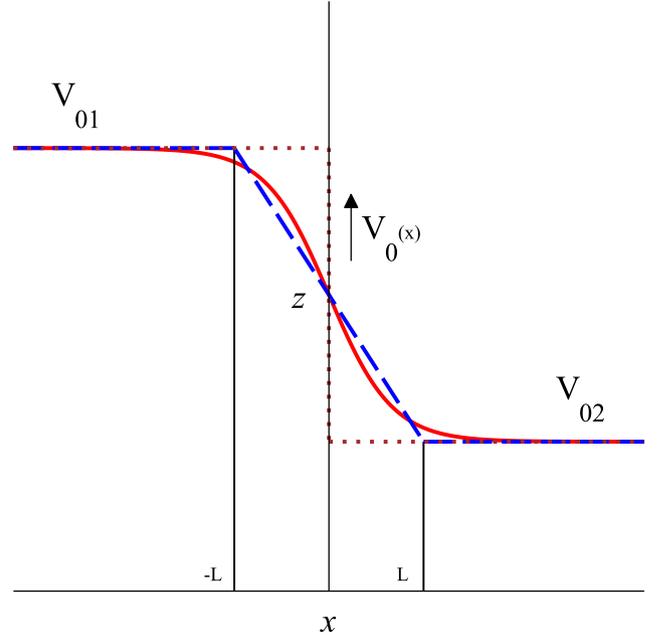

**Figure 1.** Schematic representation of shear flow along the magnetic field in the z-direction. The contact layer $2L$ wide between fast ($V_{01}$) and slow ($V_{02}$) flows has a velocity gradient along the x-axis

at $x = 0$. Here $\Delta = \frac{h-1}{h+1}$ shearing of velocity and $0 \leq \Delta \leq 1$. The dispersion equation has the form

$$\frac{\beta_A(\xi_1)}{\chi(\xi_1)} + \frac{\beta_A(\xi_2)}{\chi(\xi_2)} = 0, \quad \operatorname{Re}(\chi) \geq 0, \operatorname{Im}(\chi) \leq 0 \quad (3.1)$$

Since $\xi_1$ and $\xi_2$ differ in sign $\Delta$, then eq. (3.1) can be written as $\beta_A(\Delta)\chi(-\Delta) + \beta_A(-\Delta)\chi(\Delta) = 0$. Taking into account that the functions $\chi(\pm\Delta)$ also depend on $\Omega$, we represent this equation in the form for defining $\Omega$ as

$$\Omega^2 - 2\Delta r(\Omega)\Omega - p = 0, \quad (3.2)$$

where

$$r(\Omega) = r(\Omega; l, \Delta, \alpha, \beta, \gamma, M) = \frac{\chi(\Delta) - \chi(-\Delta)}{\chi(\Delta) + \chi(-\Delta)},$$

$$p = \frac{\alpha + \beta - 1 - M^2\Delta^2}{M^2}.$$

Note that for $l = 0$ or $\Delta = 0$, the function $r(\Omega) = 0$. In these cases, eq. (3.2) is solved exactly, $\Omega = \pm\sqrt{p}$. In the absence of a shear between velocities, $\Delta = 0$, $M^2\Omega^2 = \alpha + \beta - 1$. This means that for $\alpha + \beta < 1$ the plasma is unstable, i.e. hose instability arises $(p_\parallel > p_\perp + 2p_{mag})$. Since we are interested in KHI, in what follows we will consider the domain of parameters $\alpha + \beta > 1$, when $\Delta = 0$ the plasma is stable against small perturbations. For quasi-transverse (with respect to the flow along the magnetic field) waves, $l \to 0$, the condition for the appearance of the KHI is

$$0 < \alpha + \beta - 1 < M^2\Delta^2. \quad (3.3)$$

Consequently, with an increase in the Mach number $M$ and the shearing in velocity $\Delta$, KHI becomes stronger. The region of instability is wide, $(\alpha + \beta - 1)/M^2 < \Delta^2 \leq 1$. fig. 2 illustrates the dependence of the growth rate $\Gamma(\Delta)$ at different angles of propagation $l = \cos^2(\theta)$.

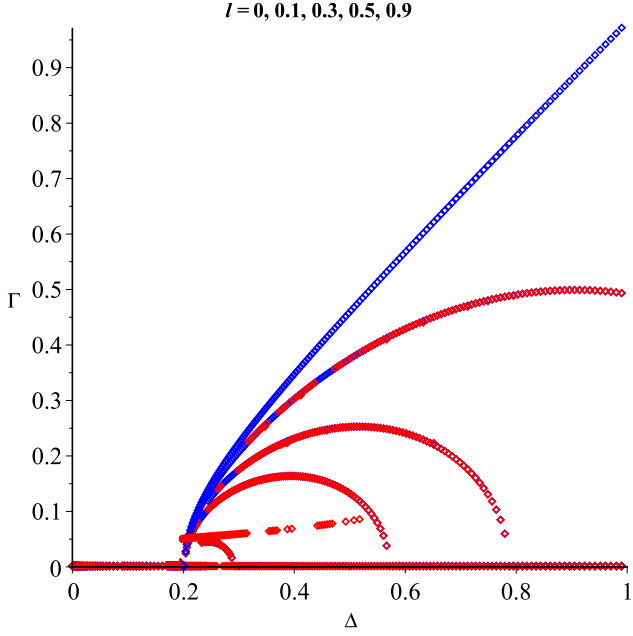

**Figure 2.** Dependence $\Gamma(\Delta)$ for different $\ell$. From top to bottom $\ell = 0, .., 0.9$, respectively. $M = 6; \alpha = 1.5; \beta = 1; \gamma = 0$;

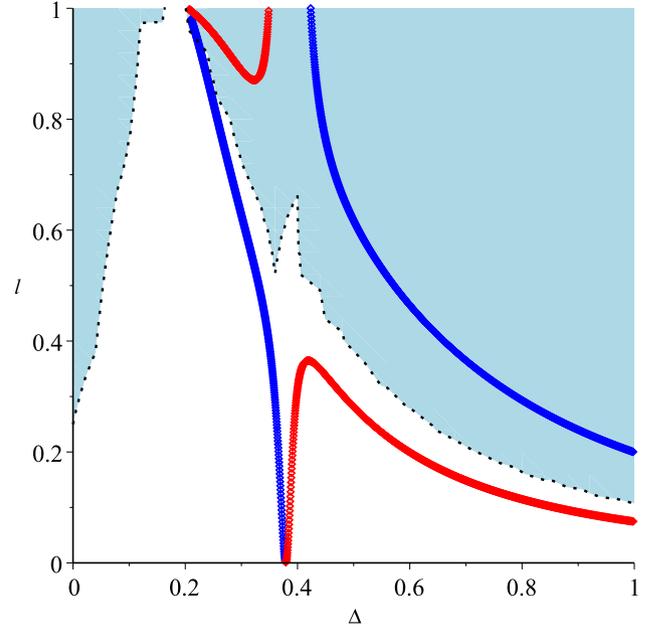

**Figure 3.** Instability threshold locations in-plane $\ell$ and $\Delta$ for $M = 6; \alpha = 1.5; \beta = 1; \gamma = 0$;. Blue region represents the condition of $F(0) < 0$.

As $l$ grows, the instability weakens; with quasi-longitudinal propagation, $l \to 1$, it disappears. Two thresholds of instability appear. The onset of instability for all $l$ is determined by condition (3.3). This condition also follows in general form from eq. (3.2) for $\Omega = 0$. For $\Omega \to 0$ this equation has a second solution. Let be

$$F(\Omega) = \frac{1}{\chi^2} = \frac{\beta_*}{(1-l)\beta_* + l\beta_A} \approx F(0) + \Omega F'(0), \quad (3.4)$$

where $F'(0) = \left.\frac{dF(\Omega)}{d\Omega}\right|_{\Omega=0}$. Substituting this expansion into eq. (3.2), we obtain that if the condition $F(0) < 0$ is satisfied, then

$$4\Delta|F(0)| + p|F'(0)| = 0 \quad (3.5)$$

The solutions of this equation determine the locations of the second instability threshold. This equation has two solutions, shown by blue and red lines in the $\Delta - l$ plane in Fig. 3 for known $\alpha, \beta, \gamma$ and $M$. In the same figure, the region of applicability to eq. (3.5) is highlighted, the condition $F(0) < 0$. Quasi-transverse waves have a wide range of instability, $0.2 < \Delta < 1$. This range narrows with increasing $l$ and the instability weakens. Dispersion eq. (3.2) depends on the plasma parameters: $\Delta, \alpha, \beta, \gamma, M$. For typical parameters of the solar wind near the Earth $V_{01} = 800$ km s$^{-1}$, $V_{02} = 400$ km s$^{-1}$, $h = 2, \Delta = 1/3, M = 6$ dependence of $\Gamma$ on $l$ for different $\beta = 0, \ldots, 5$ (magnetic fields parameter), $\alpha = 0, \ldots, 3.5$ (anisotropy parameter) and $\gamma = 0, \ldots, 2$ (heat flux parameter) are shown in Fig 4a, Fig 4b and Fig 4c, respectively. Increasing these parameters suppresses the instability. The dependence on $\gamma$ is weak.

Thus, we have shown that if the transition layer between flows has a very small width ($L \to 0$), then quasi-transverse waves become the most unstable, $l \to 0$. Let us consider the instability of these modes when the width of the transition layer is finite.

## 4 LINEAR VELOCITY PROFILE

Let us consider the linear profile of the steady flow velocity directed along the magnetic field on the $z$ axis (dashed line in Fig. 1)

$$V_0(x) = \frac{V_{01} + V_{02}}{2} - \frac{V_{01} - V_{02}}{2} \cdot \frac{x}{L}, -L \leq x \leq L. \quad (4.1)$$

Note that $V_0(-L) = V_{01}, V_0(L) = V_{02}$. For such a profile $\xi = M\left(\Omega + \Delta\frac{x}{L}\right)$. In the considered case $l \to 0$, eq. (2.9) is simplified to

$$\frac{d}{dx}\left(\beta_A(x)\frac{dY(x)}{dx}\right) - k^2\beta_A(x)Y(x) = 0, \quad (4.2)$$

or, we get differentiation with respect to $\xi$:

$$\frac{d}{d\xi}\left((p-\xi^2)\frac{dY(\xi)}{d\xi}\right) - \varepsilon^2(p-\xi^2)Y(\xi) = 0, \quad (4.3)$$

where $p = \alpha + \beta - 1, \varepsilon = \frac{kL}{M\Delta}$. Analytical solutions to this equation are expressed by the Heun Confluent Function (Decarreau et al. 1978; Ronveaux 1995; Slavyanov & Lay 2000):

$$Y(\xi) = D_1 H_1(\xi) + D_2 H_2(\xi), \quad (4.4)$$

$$H_1(\xi) = HeunC\left(0, -\frac{1}{2}, 0, -\frac{p\varepsilon^2}{4}, \frac{1+p\varepsilon^2}{4}; \frac{\xi^2}{p}\right),$$
$$H_2(\xi) = \xi\, HeunC\left(0, \frac{1}{2}, 0, -\frac{p\varepsilon^2}{4}, \frac{1+p\varepsilon^2}{4}; \frac{\xi^2}{p}\right). \quad (4.5)$$

The arbitrary constants $D_1$ and $D_2$ are determined by the boundary conditions. In case $p = 0$, solutions of eq. (4.3) are expressed by hyperbolic functions

$$H_1(\xi) = \frac{\sinh(\varepsilon\xi)}{\xi}, \quad H_2(\xi) = \frac{\cosh(\varepsilon\xi)}{\xi} \quad (4.6)$$

We introduce the notation: for $x = -L, \xi = \xi_1 = M(\Omega - \Delta)$ and in $x = +L, \xi = \xi_2 = M(\Omega + \Delta)$. Outside the region $-L \leq x \leq L$ both flows are homogeneous, $V_0 = $ const, $\xi = $ const, and in eq. (4.2) $\beta_A = $ const. Therefore, eq. (4.2) has simple, exponentially decreasing solutions:

$$Y(x) = \begin{cases} C_1 e^{k(x+L)}, & x \leqslant -L \\ C_2 e^{-k(x-L)}, & x \geq L \end{cases} \quad (4.7)$$

where $k > 0$, $C_{1,2} = $ const. Next, we apply the continuity conditions for the solutions (4.4) and (4.7) at the points $x = \pm L$ : $\{Y(x)\} = 0$, $\{Y'(x)\} = 0$, where the symbol $\{\ldots\}$ means the difference of the functions to the left and to the right of the point $x = \pm L$. We obtain

$$D_1 \left[ H_1(\xi_1) - \frac{1}{\varepsilon} H'_1(\xi_1) \right] + D_2 \left[ H_2(\xi_1) - \frac{1}{\varepsilon} H'_2(\xi_1) \right] = 0,$$
$$D_1 \left[ H_1(\xi_2) + \frac{1}{\varepsilon} H'_1(\xi_2) \right] + D_2 \left[ H_2(\xi_2) + \frac{1}{\varepsilon} H'_2(\xi_2) \right] = 0, \quad (4.8)$$

where the derivatives

$$H'(\xi_{1,2}) = \left. \frac{dH(\xi)}{d\xi} \right|_{\xi=\xi_{1,2}}.$$

The zero determinant of this system gives the desired dispersion equation

$$\left[ H_1(\xi_1) - \frac{1}{\varepsilon} H'_1(\xi_1) \right] \left[ H_2(\xi_2) + \frac{1}{\varepsilon} H'_2(\xi_2) \right] - \left[ H_1(\xi_2) + \frac{1}{\varepsilon} H'_1(\xi_2) \right] \left[ H_2(\xi_1) - \frac{1}{\varepsilon} H'_2(\xi_1) \right] = 0. \quad (4.9)$$

Substituting eq. (4.6) into eq. (4.9) yields

$$e^{2\varepsilon\xi_1} - e^{2\varepsilon\xi_2} + 2e\xi_2 e^{2\varepsilon\xi_2} - 2\varepsilon\xi_1 e^{2\varepsilon\xi_2} + 4\varepsilon^2 \xi_1 \xi_2 e^{2\varepsilon\xi_2} = 0, \quad (4.10)$$

which is valid for the special case of $p = 0$. Substituting here the values $\xi_{1,2} = M(\Omega \mp \Delta)$ and assuming that $\eta = \dfrac{2kL}{\Delta}\Omega$, we can get the exact solution (4.10) as

$$\eta^2 = (2kL - 1)^2 - \exp(-4kL). \quad (4.11)$$

Fig. 5 and 6 show the dependence of $\eta^2$ on $kL$ (the width of the transition layer normalized to the wavelength). The instability arises within the limits $0 \leq kL \leq 0.639$, where $\eta^2 < 0$. Since $\Omega = \dfrac{\Delta}{2kL}\eta$, then on passing to the tangential discontinuity ($kL \to 0$) $\eta \to 2ikL$ we obtain the previous result in (3.2): $\Omega = i\Delta$. At $\Delta \to 0$, the instability disappears.

In the general case, the solution of the dispersion eq. (4.9) with eq. (4.5) gets complicated by the HeunC functions. These functions are multi-valued complex functions, the theory of which is poorly developed. HeunC functions are represented around each singular point as infinite power (or Bessel, or hypergeometric) series. The convergence of these series near the boundary of convergence region becomes slow and requires a long computation time. For some parameter values, there is a problem with the convergence of these series. Let

$$\Omega_* = M \cdot \Omega, \quad \Delta_* = M \cdot \Delta, \quad \xi_1 = \Omega_* - \Delta_*, \quad \xi_2 = \Omega_* + \Delta_*,$$

$$k_* = k \cdot L, \quad \varepsilon = \frac{k_*}{\Delta_*}.$$

Then the solution to dispersion eq. (4.9) is function $\Omega_*(k_*, p, \Delta_*)$. Consider the case $p = 1.5$ and $\Delta_* = 2$. This corresponds to the conditions of the solar wind plasma, for example, $M = 6, \Delta = 1/3, \beta = 1, \alpha = 1.5$. The calculation results for this case are shown in Fig. 7. These figures show the complex solutions $\Omega_*(k_*)$ and, based on these solutions, the instability increments according to eq. (2.11).

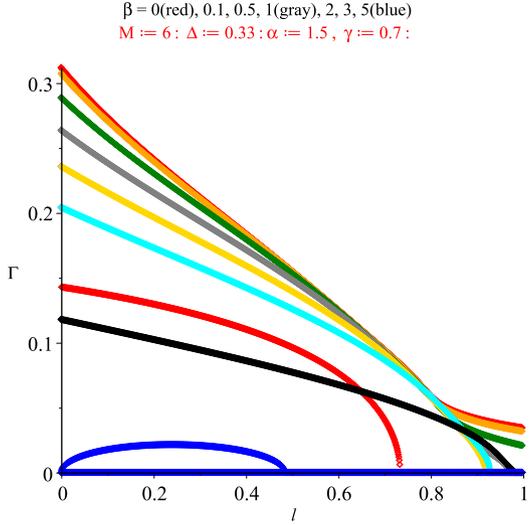

(a) From top to down $\beta = 0, .., 5$, respectively

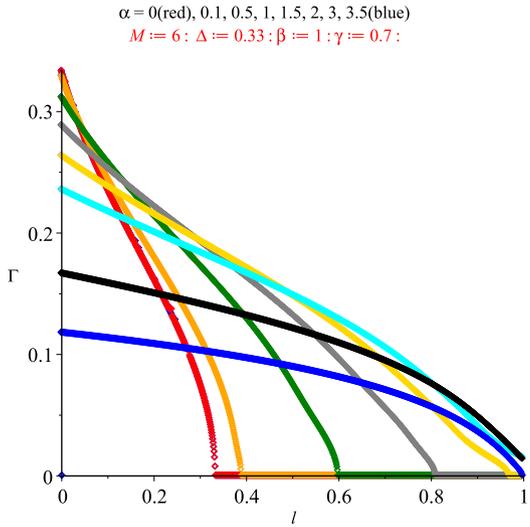

(b) From top to down $\alpha = 0, .., 5$, respectively.

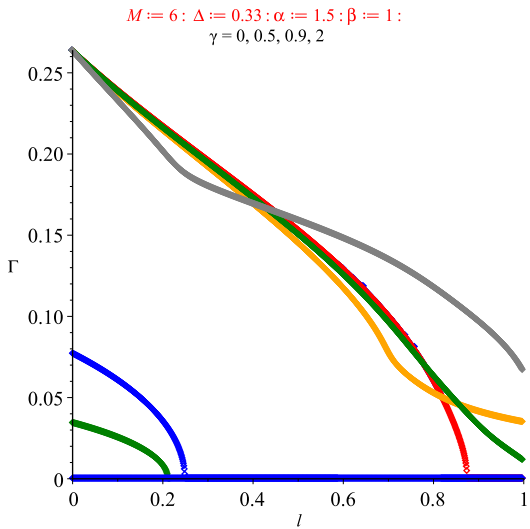

(c) From top to down $\gamma = 0, .., 5$, respectively.

**Figure 4.** The plots show the dependence between $\Gamma$ and $\ell$ for different parameters $\beta$ (a), $\alpha$ (b), and $\gamma$ (c).

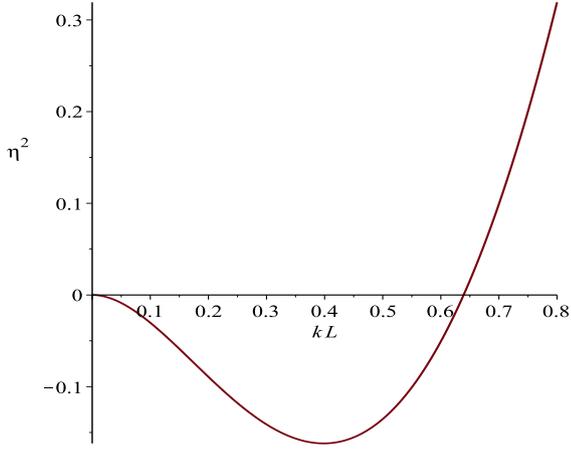

**Figure 5.** Dependence of the growth rate of instability on the width of the transition layer. Instability occurs when $\eta^2 < 0$.

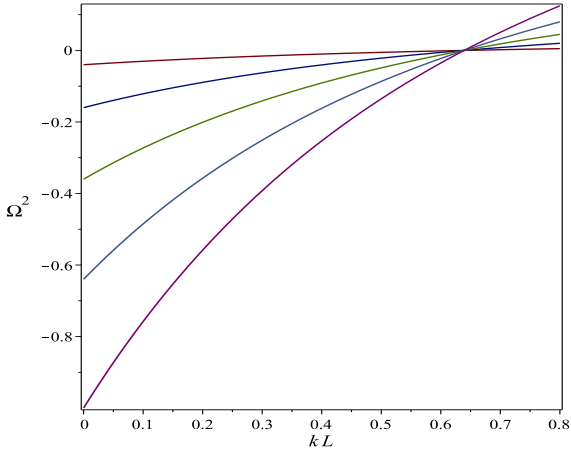

**Figure 6.** Instability condition depending on $kL$ for different $\Delta$. Curves from top to bottom correspond to $\Delta = [0.2, 0.4, 0.6, 0.8, 1.0]$

As is evident, the considered problem of a transition layer with a finite width is multi mode. All unstable modes for a given value of $k \cdot L$ have the same frequency ($Re(\omega)$), but different growth rates ($Im(\omega)$). The first mode in Fig. 7b, shown by a dashed line, is aperiodic ($Re(\omega) = 0$). The number of $n$ modes is infinite, and as $n$ grows, the instability increment increases, and the spectrum becomes continuous. Aperiodic mode is a classic KHI. With an increase in $k \cdot L = 2\pi L/\lambda$ (with an increase in the layer width $L$ or with a decrease in the wavelength $\lambda$), the instability weakens ($\Gamma$ decreases). The transition to the limit $\varepsilon = \left(\frac{kL}{M\Delta}\right) \to 0$ is exceptionally remarkable. In this case, eq. (4.3) becomes $Y'(\xi) \to \frac{1}{p-\xi^2}$. This means that the two solutions in (4.4) go to $H_1(\xi) \to \frac{1}{2\sqrt{p}} \ln\left(\frac{\xi+\sqrt{p}}{\xi-\sqrt{p}}\right)$, $H_2(\xi) \to$ const. In this limiting case, the general dispersion equation (4.9) becomes $H_1'(\xi_1) + H_1'(\xi_2) \to 0$. Therefore, $\frac{1}{p-\xi_1^2} + \frac{1}{p-\xi_2^2} \to 0$. Since $\xi_{1,2} = M(\Omega \mp \Delta)$, then we obtain that the instability condition is $M^2\Omega^2 = p - \Delta^2 M^2 < 0$, which exactly corresponds to (3.3). In Fig. 7b, the beginning of the dotted line corresponds to this limit (for the considered values of the parameters $\left(Im(\Omega_*) = \sqrt{2.5}\right)$. With an increase in $k \cdot L$ zero mode (KHI) disappears.

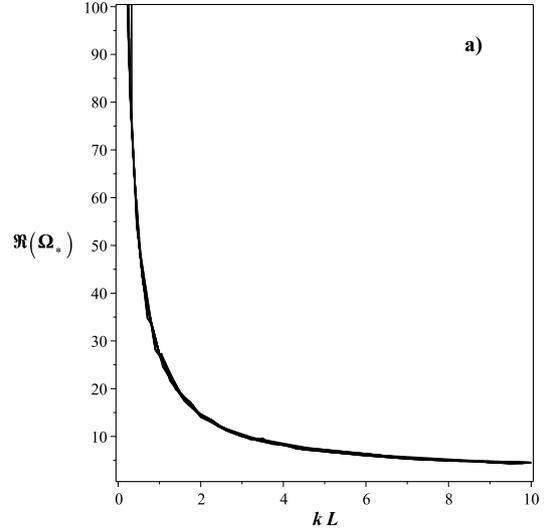
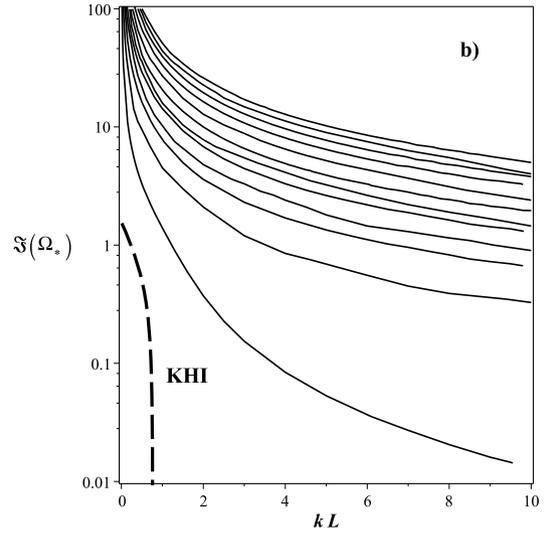
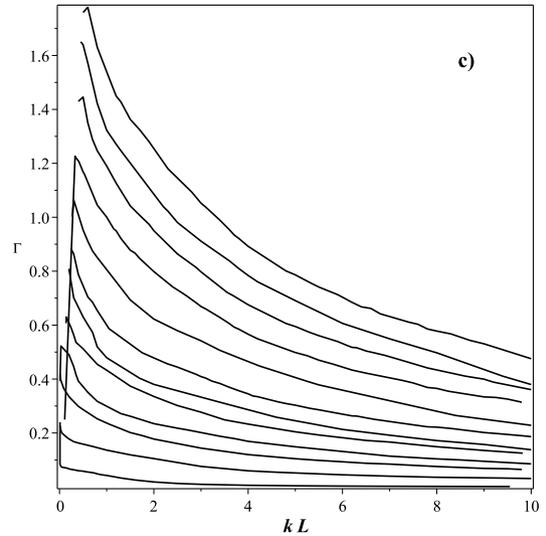

**Figure 7.** from top to down: (a) real parts of solutions, (b) imaginary parts of solutions (c) Growth rates of wave modes

## 5 HYPERBOLIC VELOCITY PROFILE

Let us now consider a hyperbolic velocity profile that smoothly describes the transition zone between flows (solid line in Fig. 1). Let be

$$V_0(x) = \frac{V_{02}e^{\sigma x} + V_{01}e^{-\sigma x}}{e^{\sigma x} + e^{-\sigma x}}, \quad (5.1)$$

where the parameter $\sigma > 0$ describes the width of the transition zone, $L \sim 1/\sigma$. Normalized velocity $V(x) = V_0(x)/\overline{V} = 1$ at $x = 0$. For such a velocity profile, the variable $\xi(x) = M(\Omega + 1 - V(x))$ varies continuously. It is convenient to enter instead of the coordinate $x$ a new independent variable $t(x) = V(x) - 1$. Therefore, $t(0) = 0, t(\pm\infty) = \mp\Delta$, i.e. ... $-\Delta \leq t \leq +\Delta$ It is easy to prove that the first derivative

$$\frac{d}{dx} = \frac{\sigma}{h-1}\left(t^2 - \Delta^2\right)\frac{d}{dt}, \quad h = \frac{V_{01}}{V_{02}} \geq 1, \Delta = \frac{h-1}{h+1}, 0 \leq \Delta \leq 1.$$

Then, bearing in mind that $\xi(t) = M(\Omega - t)$, eq. 2.9 becomes

$$\left(t^2 - \Delta^2\right)\frac{d}{dt}\left[A(t)\left(t^2 - \Delta^2\right)\frac{dY(t)}{dt}\right] - \varepsilon^2 \beta_A(t)Y(t) = 0, \quad (5.2)$$

where $\varepsilon = (h-1)(k/\sigma) > 0$. To our knowledge this equation cannot be analytically solved in general form. In case $l \to 0$ it turns into

$$\left(t^2 - \Delta^2\right)\frac{d}{dt}\left[\left(p - (\Omega - t)^2\right)\left(t^2 - \Delta^2\right)\frac{dY(t)}{dt}\right] - \varepsilon^2\left[p - (\Omega - t)^2\right]Y(t) = 0, \quad (5.3)$$

where $p = (\alpha + \beta - 1)/M$. Eq. (5.3) remains difficult for analytical study due to singularities. However, in the special but important case $p = 0$ this equation can be reduced to the Heun equation (Decarreau et al. 1978; Ronveaux 1995; Slavyanov & Lay 2000), the solutions of which are

$$Y(t) = \frac{1}{\Omega - t}\left[C_1 H_1(t)\left(\frac{\Delta + t}{\Delta - t}\right)^{\frac{\varepsilon}{2\Delta}} + C_2 H_2(t)\left(\Delta^2 - t^2\right)^{-\frac{\varepsilon}{2\Delta}}\right]. \quad (5.4)$$

Here $C_1$ and $C_2$ are arbitrary integration constants that must be determined from the boundary conditions, and $H_1(t)$ and $H_2(t)$ are Heun functions:

$$H_1(t) = HeunG\left(\frac{2\Delta}{\Delta + \Omega}, -\frac{2\Delta}{\Delta + \Omega}, -1, 2, 1 + \frac{\varepsilon}{\Delta}, 0; \frac{t + \Delta}{\Delta + \Omega}\right),$$

$$H_2(t) = HeunG\left(\frac{2\Delta}{\Delta + \Omega}, \frac{\varepsilon(\Delta + \Omega)(\varepsilon - \Delta) - 2\Delta^3}{(\Delta + \Omega)\Delta^2}, 2 - \frac{\varepsilon}{\Delta},\right. \quad (5.5)$$

$$\left. -1 - \frac{\varepsilon}{\Delta}, 1 - \frac{\varepsilon}{\Delta}, 0; \frac{t + \Delta}{\Delta + \Omega}\right).$$

Solutions (5.4) must satisfy the boundary conditions $|Y(t)|_{t=\pm\Delta} = 0$. Since $H_{1,2}(-\Delta) = 1$, then we obtain that $C_2 \equiv 0$. Another condition for the boundedness of the solution for $t = +\Delta$ will be $H_1(+\Delta) = 0$. This condition is the dispersion equation for determining $\Omega(k)$:

$$HeunG(\psi, -\psi, -1, 2, 1 + s, 0; \psi) = 0, \quad (5.6)$$

where $\psi = \frac{2\Delta}{\Omega + \Delta}$ and $s = \frac{\varepsilon}{\Delta} = (h + 1)\frac{k}{\sigma}$. Using the properties of Heun's function, we can represent the desired dispersion equation in the form

$$HeunG(\phi, -1, -1, 2, 1 + s, 1 - s, 1) = 0, \phi = \frac{1}{\psi} = \frac{1}{2} + \frac{\Omega}{2\Delta}. \quad (5.7)$$

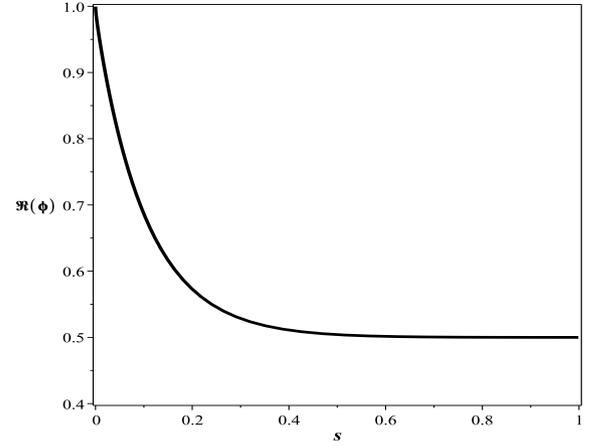

**Figure 8.** The real part of the dispersion equation's solutions (5.7)

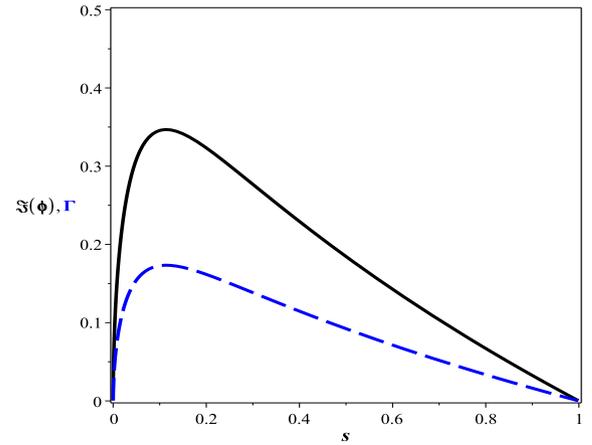

**Figure 9.** The imaginary part of the solutions of the dispersion equation (5.7) (black solid line) and the growth rate of instability (dashed line).

This equation determines the complex dependence $\phi(s)$. For the instability threshold ($\Omega = 0$), the parameter $\phi = \frac{1}{2}$ and the Heun function transforms into the hypergeometric Gaussian function

$$HeunG\left(\frac{1}{2}, -1, -1, 2, 1 + s, 1 - s, 1\right) = {}_2F_1(-1, 2; 1 + s, 1) = \frac{s - 1}{s + 1} = 0. \quad (5.8)$$

Thus, if $s = (h + 1)\frac{k}{\sigma} = 1$, then $\Omega = 0$. Numerical solutions (5.7) are shown in Fig. 8 and 9. Instability growth rate (2.11) is defined as

$$\Gamma = \frac{2\Delta \operatorname{Im}(\phi)}{1 + \Delta[2\operatorname{Re}(\phi) - 1]}. \quad (5.9)$$

The found solutions is shown in Fig. 9 with a dotted line for the case $h = 2$. Instability is possible in the region $kL \approx s < 1$.

The eigenfunctions for the found complex eigenvalues $\{s, \phi\}$ for a given $\Delta$ can be calculated by the formula

$$Y(t) = \frac{C_1}{\Omega - t} \cdot \left(\frac{\Delta + t}{\Delta - t}\right)^{\frac{s}{2}} \cdot HeunG\left(\phi, -1, -1, 2, 1 + s, 1 - s, \frac{t + \Delta}{2 \cdot \Delta}\right) \quad (5.10)$$

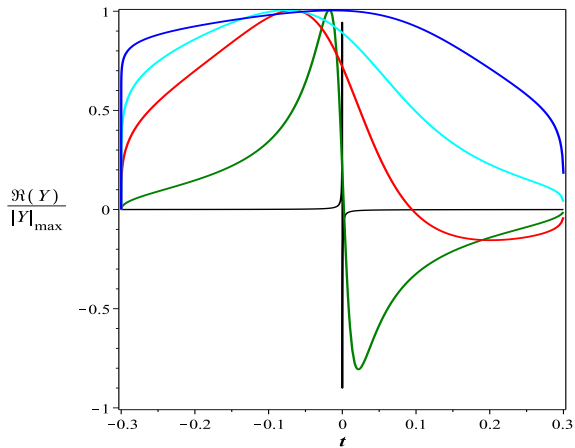

**Figure 10.** Eigenfunctions for given eigenvalues for $s$ = 0.1 (blue), 0.3 (cyan), 0.5 (red), 0.9 (green), 0.9995 (black)

where the frequencies are defined as $\Omega := 2 \cdot \Delta \cdot \left(\phi - \frac{1}{2}\right)$. Fig.10 shows the case $\Delta = 0.3$ for the normed eigenfunctions. Due to the uncertainty of the constant $C_1$, the eigenfunctions are normalized to the corresponding maxima

{$s, |Re(Y)|_{max}$} = [0.1, 1.9633; 0.3, 2.3641; 0.5, 3.0828; 0.9, 13.48;

0.9995, 4369.45]

In the region $t = [-\Delta, +\Delta]$, abrupt changes in functions in short distances occur at $t \to 0$ and $s \to 1$. For long waves, when $s < 1$, substantial changes occur on a large scale.

## 6 CONCLUSIONS

There are various cosmic situations where a hot plasma that is very rarefied (weakly collisional) and has a strong magnetic field can become significantly anisotropic concerning the direction of the magnetic field. Consequently, a heat flux occurs along the magnetic field line in this scenario. Under such circumstances, novel varieties of MHD instability in plasma emerge - mirror, fire hose, and other analogs from low-frequency kinetics (e.g. Kuznetsov & Dzhalilov 2009; Dzhalilov et al. 2011) - as well as other types of MHD instabilities. The plasma produced by the solar wind is an excellent illustration of this type of plasma in action. In addition, the anisotropy of this plasma has previously been established in many space experiments. It is also proven that there is low-frequency large-scale Alfén turbulence in the solar wind plasma, which can be observed (Coleman 1967, 1968; Belcher et al. 1969).

Low-frequency Alfén waves play an essential role in accelerating particles, the solar wind's occurrence, and the supersonic wind's asset in the heliosphere. However, it is not easy to comprehend the dissipation of Alfén waves and the heating of plasma in the conditions of the solar corona and the solar wind. To be dissipated the large-scale Alfén turbulence through a nonlinear cascade, a primary basic mechanism is required. As a result of the operation of such a mechanism, the plasma should become structured. Viall et al. (2021) review observations of periodic density structure moving outward from the corona into the inner Heliosphere. Periodic density oscillations are typically found in the slow solar wind with 4–140 min periods, corresponding to radial wavelengths of $8 \times 10^4 - 3 \times 10^6$ km

(Viall et al. 2008; Kepko et al. 2020). In addition, proton number-density periodic structures are seen to persist in the solar wind well past 1 AU (Birch & Hargreaves 2020).

It is known that the solar wind has a bi-modal structure, and due to rotation, SIR (a solar wind stream interaction region) arises. Under such conditions, shear flows in the interfaces, and related instabilities naturally arise. These instabilities can transfer the energies of large-scale Alfvén oscillations into small-scale motions and dissipate them. Kelvin–Helmholtz instability (KHI) is a fundamental physical process in fluids and magnetized plasmas, observed at boundaries of flows in different astrophysical and geophysical situations. MHD KHI has been well studied under conditions of isotropic collisional plasma. The solar wind conditions are anisotropic, and there is a heat flow. In this paper, we set the goal of studying the properties of KHI under SIR conditions observed in the Earth's orbit. Based on 16-moment MHD equations derived for the moments of the kinetic equations in the low-frequency regime, a general equation is obtained that describes the KHI at the interface of two plasma flows with anisotropic properties. We considered identical anisotropic plasmas inflows but with different supersonic velocities ($M = 6$) to simplify the problem. We studied in detail the case when the waves are directed almost perpendicular to the magnetic field and the flow. In this case, it was possible to analytically solve problems when the transition region between flows has a finite width, a necessary realistic condition. The plasma parameters are chosen so that in the initial state, the plasma is stable regarding mirror and firehose instabilities, i.e., the coupling of KHI by other global modes is excluded in this work. It has been discovered that the conditions for the layer width concerning the wavelength under which the occurrence of KHI diminishes can be satisfied. In contrast to a tangential discontinuity (when the transition layer width is zero), the growing rate of KHI can increase arbitrarily with the wavenumber $k$, but for a finite width $L$, there is $k_{max}$ greater of which KHI cannot occur. The values of $k_{max}$ generally depend on the plasma parameters.

## ACKNOWLEDGEMENTS

This work was supported by the Science Development Foundation under the President of the Republic of Azerbaijan - Grant № EİF-BGM-4-RFTF-1/2017-21/06/1 and work of R.I. was supported by the European Research Council (ERC) under the European Union's Horizon 2020 research and innovation programme (grant agreement No 724326).

## DATA AVAILABILITY

The data underlying this article will be shared on reasonable request to the corresponding author.

This paper has been typeset from a TeX/LaTeX file prepared by the author.